\def\be{\begin{equation}}
\def\ee{\end{equation}}
\def\kt{k_{\perp}}
\def\pt{p_{\perp}}

\def\<{\langle}
\def\>{\rangle}

\newcommand{\bea}{\begin{eqnarray}}
\newcommand{\eea}{\end{eqnarray}}
\newcommand{\nn}{\nonumber\\}

      [1999/12/01 v1.4c Il Nuovo Cimento]
\documentclass{cimento}
\usepackage{amsmath,amssymb,amsfonts,graphicx}
\usepackage{bm}
\usepackage{slashed}
\usepackage[caption=false]{subfig}
\usepackage[svgnames]{xcolor}

\title{Fully Differential Monte-Carlo Generator Dedicated to TMDs and Bessel-Weighted Asymmetries}
\author{M.~Aghasyan\from{ins:x}\ETC,
 H.~Avakian\from{ins:y}}
\instlist{\inst{ins:x} INFN, Laboratori Nazionali di Frascati, 00044 Frascati, Italy
\inst{ins:y} JLab, 12000 Jefferson Ave, Newport News, VA 23606, USA}

\PACSes{\PACSit{13.60.-r,  13.87.Fh, 13.88.+e}
\PACSit{14.20.Dh,  24.85.+p}{}}
\begin{document}

\maketitle

\begin{abstract}
We present studies of double longitudinal 
spin asymmetries in semi-inclusive deep inelastic 
scattering  using a new dedicated Monte Carlo generator, which includes quark intrinsic transverse momentum
 within the generalized parton model 
based on the fully differential cross section for the process. 
Additionally, we apply Bessel-weighting to the 
simulated events to extract transverse momentum dependent 
parton distribution functions 
and also discuss possible uncertainties due to kinematic correlation effects. 
\end{abstract}

\section{Fully differential SIDIS cross section}
The transverse momentum dependent (TMD) partonic distribution and fragmentation functions play a crucial role in measuring and interpreting information towards a true 3-dimensional imaging of the nucleons. TMD PDFs can be accessed in several experiments, but the main source of information is semi-inclusive deep inelastic scattering (SIDIS) of polarized leptons off polarized nucleon. 
For SIDIS,  the theoretical formalism is described in a series of papers~\cite{PhysRevD.71.074006,PhysRevD.84.034033} using a tree level factorization~\cite{Mulders:1995dh} where  the standard momentum convolution integral~\cite{Bacchetta:2006tn} relates the quark
intrinsic transverse momentum in a nucleon to the transverse momentum of the produced hadron $P_{hT}$.

In this work we present a model independent extraction of the ratio of polarized, $g_{1L}$, and unpolarized, $f_1$, TMD distributions using
a Monte Carlo (MC) generator based on the fully differential
cross section, in which we reconstruct the transverse momentum of the final hadron
after MC integration over the quark intrinsic transverse momentum. 
In the MC generator we used the SIDIS cross section described in Ref. \cite{PhysRevD.71.074006}. 
The Bessel-weighted asymmetry, providing access to the ratio of Fourier transforms of $g_{1L}$ and  $f_1$  \cite{Boer:2011xd}, 
has been extracted. The uncertainty of the extracted TMDs was estimated using unintegrated, transverse momentum {\it non}-factorized distribution and fragmentation functions.

The MC generator software employs the general-purpose, self-adapting MC event generator Foam \cite{TFOAM} for drawing random points according to an arbitrary, user-defined distribution in the $n$-dimensional space.

We consider the following SIDIS process
\be
{\ell}(l) + N(P)\rightarrow \ell(l') + h(P_{h}) + X, 
\ee
were $\ell$ -is the scattered lepton, $N$ is the proton target and $h$-is the observed hadron (four-momenta notations given in parentheses). The virtual photon momentum is defined $q = l-l'$ and its virtuality $Q^2=-q^2$. Following the Trento conventions \cite{Bacchetta:2004jz}, we use the virtual photon-nucleon center of mass system, where the virtual photon momentum $q$ is along the $z$ axis and the proton momentum $P$ is in the opposite direction. The detected hadron $h$ has momentum $P_h$. In the parton model the virtual photon scatters off an on-shell quark. The initial quark momentum $k$ and scattered quark momentum $k'$ have the same intrinsic transverse momentum component $\kt$ with respect to the $z$ axis. The initial quark has  a fraction $x$ of the proton's light-cone momentum, while the produced hadron  $h$ has a light-cone momentum fraction $z$ and transverse momenta component $\pt$ with respect to the scattered quark's momentum $k'$ (see Ref.~\cite{PhysRevD.71.074006}).
The fully differential cross section used in the MC generator is then given by \cite{PhysRevD.71.074006}:
\bea
\frac{d \sigma}{ dx dy dzd^2 {\bf p}_{\perp}  d^2 {\bf k}_{\perp} } &=& K(x,y)J(x,Q^2,\kt)\times \nn 
&& \hspace{ -0.5cm} \times \left[  \sum_q f_{1,q}(x,k_{\perp})D_{1,q}(z,p_{\perp})  + \lambda \sqrt{1-\varepsilon^2}g_{1L,q}(x,k_{\perp})D_{1,q}(z,\pt)  \right] 
\label{FDALL}
\eea
where the summation runs over the quark flavors in the target.  The kinematic factors $K(x,y)$ and $\varepsilon$, and the Jacobian $J(x,Q^2,\kt)$ are defined in \cite{PhysRevD.71.074006}. $\lambda$ represents the product of lepton and nucleon helicities ($\lambda=\pm1$). $ f_{1,q}(x,k_{\perp})$ and $g_{1L,q}(x,k_{\perp})$ are the quark TMD distributions and $D_{1,q}(z,\pt)$ is the TMD fragmentation function for an unpolarized quark $q$.

\begin{figure}[hbt]
\centerline{\includegraphics[height=3.5in]{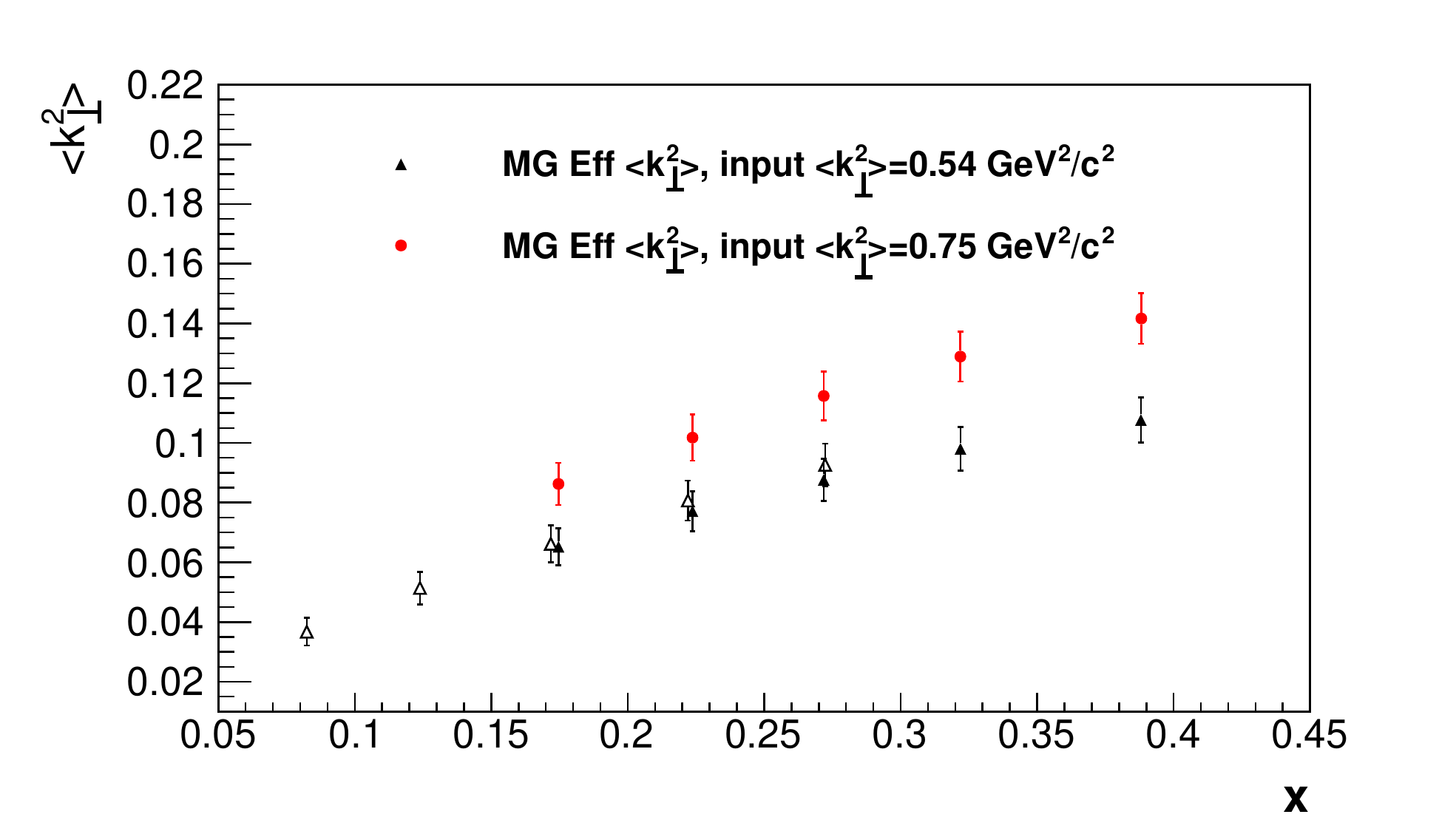}}
\caption{\scriptsize (Color online) The effective $\<k^2_{\perp}(x)\>$ in the modified Gaussian (MG) ansatz
for the TMD distributions (see Eq.~\ref{BrodskyDF}) as function of $x$ for
two different input parameters $<k^2_{\perp}\>$
obtained from MC events simulated within the range $0.50<z<0.52$ and using
$6~{\rm GeV}$ (solid symbols) and $11~{\rm GeV}$ (empty symbols) lepton beam energies.}
\label{kTvsxB}
\end{figure}

In many  phenomenological studies of the semi-inclusive deep inelastic scattering, the transverse momentum dependence of distribution and fragmentation functions is factorized from the light-cone momentum dependence $x$ and $z$, respectively. In our MC generator we use the Bessel-weighting method~\cite{Boer:2011xd} to alleviate this approximations in the extraction of TMD distributions from SIDIS measurements. 
We discuss the following modified Gaussian (MG) expressions for the TMD distributions and fragmentation function,
inspired by AdS/QCD \cite{SJBrodskyWarsaw2012,PhysRevLett.102.081601}, in which $x$ and $\kt$ ($z$ and $\pt$)
are not factorized:
\be
f_1(x,\kt) =f_1(x) \frac{e^{-\frac{\kt^2}{x(1-x)\<\kt^2\>_{f_{1}}}}}{x(1-x)\<\kt^2\>_{f_{1}}}, ~g_{1L}(x,\kt) =g_{1L}(x) \frac{e^{-\frac{\kt^2}{x(1-x)\<\kt^2\>_{g_{1}}}}}{x(1-x)\<\kt^2\>_{g_{1}}}
\label{BrodskyDF}
\ee
\be
D_1(z,\pt) = D_1(z) \frac{e^{-\frac{\pt^2}{z(1-z)\<\pt^2\>}}}{z(1-z)\<\pt^2\>}
\label{BrodskyFF}
\ee

TMD distributions non-factorized in $x$ and $\kt$ are also suggested by the diquark spectator model \cite{PhysRevD.77.094016} and the NJL model \cite{Matevosyan:2011vj}.  
For a test extraction, the $x$ and $z$ dependence in Eqs.~(\ref{BrodskyDF}-\ref{BrodskyFF})
we use the parametrizations $f_1(x) = (1-x)^3\, x^{-1.313}$, $g_{1L}(x) = f_1(x)x^{0.7}$ and $D_1(z) = 0.8\, (1-z)^2$,
with widths: $\<\kt^2\>_{f_{1}}= 0.54 \,{\rm GeV}^2$ \cite{SJBrodskyWarsaw2012}, $\<k^2_{\perp}\>_{g_{1}} = 0.8\<k^2_{\perp}\>_{f_{1}}$ and $\<\pt^2\>=0.14 \,{\rm GeV}^2$. 
Fig.~\ref{kTvsxB} illustrates the average effective transverse momentum $\<\kt^2\>= \<\kt^2\>_{f_{1}}(1-\<x\>)\<x\>$ 
as function of $x$ from MC events for $6~{\rm GeV}$ (solid symbols) and $11~{\rm GeV}$ (empty symbols) 
incoming electron beam energies and produced hadrons within $0.5<z<0.52$. 
Note that the value for $\<\kt^2\>_{f_{1}}$ obtained from the MC events is always smaller than the implemented 
value due to energy and momentum conservation \cite{Mupcome}. 
A similar non-flat dependence of the average quark transverse momenta versus $x$ and $z$ is also observed in NJL-jet model \cite{Matevosyan:2011vj}.

\section{Bessel-weighted extraction of the double spin asymmetry $A_{LL}$}

We present the extraction of the double spin asymmetry $A_{LL}$, defined as the ratio of the difference and the sum of electroproduction cross sections for antiparallel, $\sigma^+$, and parallel, $\sigma^-$, configurations of lepton and nucleon spins, using the Bessel-weighting procedure described in \cite{Boer:2011xd} and applied in \cite{Lu:2012ez}.
Within this approach, one can extract the Fourier transform of the double spin asymmetry, 
$ A^{J_{0}(b_TP_{hT})}_{LL}(b_T)$, defined as
\be
 A^{J_{0}(b_TP_{hT})}_{LL}(b_T) = \frac{ \tilde \sigma^+(b_T) - \tilde \sigma^-(b_T)}{\tilde \sigma^+(b_T) + \tilde \sigma^-(b_T) }=\frac{\tilde \sigma_{LL}(b_T)}{\tilde \sigma_{UU}(b_T)}=\sqrt{1-\varepsilon^2} \frac{\sum_{q}\tilde g^{q}_1(x,z^2b_T^2) \tilde D^{q}_{1}(z,b_T^2)}{\sum_{q}\tilde f^{q}_1(x,z^2b_T^2) \tilde D^{q}_{1}(z,b_T^2)},   
 \label{tildall}
\ee 
 using measured double spin asymmetries as functions of $P_{hT}$~\cite{Avakian:2010ae},  for fixed $x$, $y$, and $z$ bins.
Here $b_T$ is the Fourier conjugate of the $P_{hT}$. The Fourier transforms of the helicity dependent cross sections, $\sigma^{\pm}( b_T)$, can be extracted by integration (analytic models) or summation (for data and MC) over the hadronic transverse momentum, weighted 
by a Bessel function $J_{0}$,
\be
 \tilde \sigma^{\pm}( b_T) \simeq S^{\pm}=\sum_{i=1}^{N^{\pm}} J_{0}(b_T P_{hT,i}) \,  .
 \label{spm}
\ee

\begin{figure}[hbt]
\centerline{\includegraphics[height=3.5in]{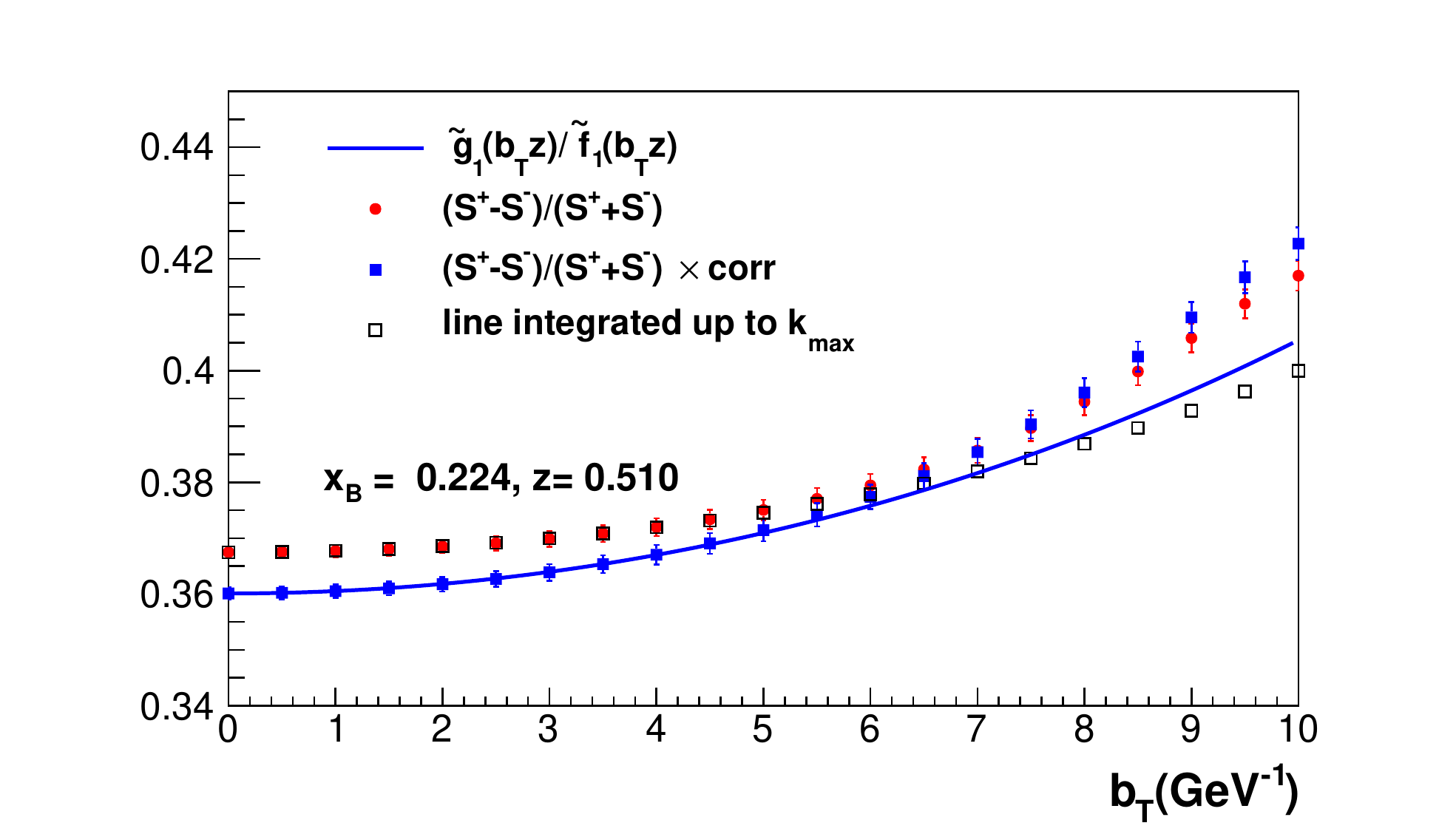}}
\caption{\scriptsize (Color online) Bessel-weighted asymmetry vs $b_T$ with and without the correction, together with analytical and numerical comparison from the MC with PDFs and FFs from Eq.~\ref{BrodskyDF}. See the text for more details. }
\label{BWPhTcorr}
\end{figure}

The Bessel-weighted asymmetry obtained from the simulated events is shown in Fig.~\ref{BWPhTcorr} as
function of $b_T$ with 
filled (red) circles, while the analytic
expression $\frac{\tilde g_1(x,zb_T)}{\tilde f_1(x,zb_T)}$ using $\<\kt^2\>_{g_1}$ and $\<\kt^2\>_{f_1}$ 
from the fits to $\kt^2$ distributions from the same MC sample 
is depicted by the (blue) full line.
For values $b_T < 6~{\rm GeV}^{-1}$, which corresponds to about $1~fm$,  the  Bessel-weighted asymmetries could 
be extracted with an accuracy of 2.5\%, although with a systematic shift. 
This clear systematic shift between the extracted and calculated asymmetries
is due to the  kinematic restrictions introduced by energy and momentum conservation\footnote{In the light-cone coordinate system the transverse component is less or equal to the momentum component along the light-cone vector.}
as well as binning effects, which deform the Gaussian shapes 
of the $\kt$ and $\pt$ distributions. 
In experiments, there is always a cutoff at high $P_{hT}$ due to acceptance and the small cross section,  
as well as a cutoff at small $P_{hT}$ where the azimuthal angles are not well defined due to the
experimental resolution. 
These restrictions in $P_{hT}$ directly affect the extracted $\kt$ and $\pt$ distributions and yield to the
mentioned distortion of Gaussian shapes which result in the systematic shift between the
extracted and the calculated asymmetries. 
Obviously, this shift depends on experimentally introduced restrictions for the accessible $P_{hT}$ range. 

We discuss two approaches which take these conditions into account, one corrects the data (simulated
MC events) the other applies limits on the integration range
for the intrinsic transverse parton momenta when calculating the asymmetry. 
In order to correct the data, the contributions from the missing $P_{hT}$ ranges outside the accessible 
values is estimated by choosing a certain model for the parton transverse momentum dependence, e.g. a 
Gaussian distribution.
This model dependent correction of data is shown in Fig.~\ref{BWPhTcorr} by the (blue) filled squares. 
The extracted asymmetry does now match the theoretical curve for values $b_T < 6~{\rm GeV}^{-1}$.
Alternatively, 
limited numerical integration over intrinsic transverse momenta in the calculation of the 
asymmetry, where the integration limits correspond to the accessible experimental $P_{hT}$ range, 
yields a calculated asymmetry that describes correctly the experimental situation without introducing 
a model dependence. 
This is shown in Fig.~\ref{BWPhTcorr} by the (black) open squares.



In summary, we used a fully differential Monte Carlo event generator to study 
how the implemented quark transverse momentum distributions $\kt$ and $\pt$ 
are being changed due to the energy and momentum conservation low and kinematic correlations within 
the generalized parton model. 
We then use the obtained $\kt$ and $\pt$ distributions to apply the Bessel-weighting strategy 
for the extraction of TMD distributions. 
As example we calculate the Bessel-weighted double spin asymmetry $A^{J_{0}(b_TP_{hT})}_{LL}(b_T)$.
Within the $b_T$ range of $b_T< 6~{\rm GeV}^{-1}$, which corresponds to about $1~fm$,  
the Bessel-weighted asymmetries could be extracted within 2.5\% accuracy. 

\acknowledgments
We thank M. Anselmino, U. D'Alesio, S. Melis, L. Gamberg, A. Prokudin, A. Kotzinian, S. Brodsky, B. Musch, D. Boer and  H. Matevosyan for useful and stimulating discussions.
This work was supported by the U.S. Department of Energy
and the National  Science Foundation, 
the Italian Istituto Nazionale di Fisica Nucleare, by the Research Infrastructure Integrating Activity
Study of Strongly Interacting Matter (acronym HadronPhysic3, Grant Agreement
n. 283286) under the Seventh Framework Programme of the European Community.

\bibliography{Aghasyan_QCDN12}

\end{document}